\documentclass[12pt]{article}
\pdfoutput=1
\usepackage{epsf,amsfonts,amssymb,epsfig,amsmath,graphics,slashed}
\usepackage{hyperref,hep,graphicx,subfig}

\newcommand{\nn}{\notag \\}

\begin{document}

\makeatletter
\renewcommand{\theequation}{\thesection.\arabic{equation}}
\@addtoreset{equation}{section}
\makeatother

\baselineskip 18pt

\begin{titlepage}

\vfill

\begin{flushright}
Imperial/TP/2012/JG/02\\
\end{flushright}

\vfill

\begin{center}
   \baselineskip=16pt
   {\Large\bf Black holes dual to helical current phases}
  \vskip 1.5cm
      Aristomenis Donos and Jerome P. Gauntlett\\
   \vskip .6cm
      \begin{small}
      \textit{Blackett Laboratory, 
        Imperial College\\ London, SW7 2AZ, U.K.}
        \end{small}\\*[.6cm]

\end{center}

\vfill

\begin{center}
\textbf{Abstract}
\end{center}

\begin{quote}
We consider the class of $d=4$ CFTs at finite temperature and chemical potential that are holographically described within
$D=5$ Einstein-Maxwell theory with a Chern-Simons term. 
The high temperature phase, which is spatially homogeneous and isotropic, is dual to the AdS-Reissner-N\"ordstrom black brane solution. For sufficiently 
large Chern-Simons coupling, we construct new electrically charged AdS black hole solutions that are dual to the low temperature, spatially modulated phase. In this phase the current, associated with the abelian global symmetry, spontaneously acquires a helical order. The new
black holes are stationary and also have Bianchi VII$_0$ symmetry.

\end{quote}

\vfill

\end{titlepage}
\setcounter{equation}{0}


\section{Introduction}
Spatially modulated phases, in which the Euclidean spatial symmetry is spontaneously broken down to some smaller subgroup,
appear in condensed matter systems in a wide variety of settings, including spin density waves \cite{Gruner:1994zz}, charge density waves \cite{Gruner:1988zz} and FFLO states \cite{larkin:1964zz,Fulde:1964zz},
and are also anticipated in QCD at high baryonic density \cite{Deryagin:1992rw}. It is therefore of
interest to investigate the properties of such phases for strongly coupled matter using the AdS/CFT correspondence. In this context it has been argued that
spatially modulated phases can arise for CFTs at finite temperature and charge density \cite{Nakamura:2009tf,Ooguri:2010kt,Donos:2011bh,Takeuchi:2011uk,Donos:2011ff,Donos:2012gg} and also in the presence
of magnetic fields \cite{Donos:2011qt,Donos:2011pn}. 
Other work, utilising the brane probe approximation, can be found in 
\cite{Domokos:2007kt,Ooguri:2010xs,Bayona:2011ab,Bergman:2011rf}. The simplicity of these constructions has
led to the speculation that the typical ground states of holographic matter at finite charge density and/or in a magnetic field could be spatially modulated \cite{Donos:2011ff}.

The first construction of black holes dual to spatially modulated phases was recently presented in the context of a $D=5$ gravitational model with a gauge field and a charged two-form in \cite{Donos:2012gg}. 
Electrically charged AdS$_5$ black holes were constructed which are holographically dual to $p$-wave superconductors with a helical order. The black holes of \cite{Donos:2012gg} are static and have a Bianchi VII$_0$ symmetry, which is naturally associated with the helical order. It was also shown that at zero temperature the black holes
become smooth domain wall solutions interpolating between $AdS_5$ in the UV and a homogeneous but non isotropic ground state with a scaling symmetry in the IR, of a type similar to those found in \cite{Iizuka:2012iv}.

Here we will consider another class of $D=5$ models, first studied in this context in \cite{Nakamura:2009tf}, namely Einstein-Maxwell theory with a Chern-Simons term. This class of models, parametrised by the strength of the Chern-Simons coupling, $\gamma$, can be used to study $d=4$ CFTs with an abelian 
global symmetry, whose anomaly is fixed by $\gamma$. We will be interested in phases of the dual CFT at finite temperature and chemical potential with respect to the global symmetry which means that we need to construct electrically charged AdS$_5$ black holes. At high temperatures the CFTs are described
by the standard AdS-Reissner-N\"ordstrom black brane solution. When $\gamma$ is greater than a specific critical value, $\gamma_c$, the AdS-RN black brane has spatially modulated instabilities below a critical temperature, suggesting that the system moves into a spatially
modulated phase in which the current acquires a helical order \cite{Nakamura:2009tf}. In the limit $\gamma\to\infty$, where the back-reaction to gravity gets switched off, it was argued that the phase transition should be second order with mean field behaviour \cite{Ooguri:2010kt}.

The purpose of this paper is to construct the fully back reacted spatially modulated black holes for $\gamma>\gamma_c$. As in \cite{Donos:2012gg}, the key ingredient in the construction of the new black hole solutions is that they have a Bianchi VII$_0$ symmetry associated with the helical order. With this symmetry, combined with time-translation invariance, we can use an ansatz involving several  functions which just depend on a single radial coordinate.
After substituting into the equations of motion we are led to a system of ODE's which we solve numerically.
While the black holes in \cite{Donos:2012gg} were
static here they are just stationary and this leads to the dual phase having spatially modulated momentum in addition to the spatially modulated pressure and shear that was seen in \cite{Donos:2012gg}. The new black hole solutions exist for temperatures lower than the critical temperature (which is not always the case \cite{Buchel:2009ge}) and have less free energy than the AdS-RN black branch. The helical current phase is thus thermodynamically preferred and we show that it is always second order with mean field
behaviour. We will see that the spatial modulation persists in the $T\to 0$ limit and in this limit the entropy density approaches zero.

Our black holes are dual to helical current phases which are reminiscent of the ``chiral nematic (or ``cholosteric") 
phase of liquid crystals (e.g. \cite{dejennes}). Recall that the order parameter for a nematic phase is a three-dimensional unit vector ${\bf n}$,
defined up to sign, called the ``director". 
In the chiral nematic phase there is a helical structure in which the director twists along an axis perpendicular to the direction of the director. For wave-number $k$ the pitch of a helical phase is $p=2\pi/k$.
While a general helical phase, including ours, is periodic with period $p$, for a chiral nematic it has period $p/2$
because ${\bf n}\cong-{\bf n}$. One important issue is to calculate the temperature dependence of the pitch of the helical order (e.g. \cite{jacksonsz}). In many materials the pitch increases with decreasing temperature but materials are known for which it decreases. We can obtain the precise temperature dependence of the pitch for our helical phases, finding that it monotonically increases, approaching a non-zero value at $T=0$.
Chiral nematics are well known to have interesting optical properties\footnote{This was recently explored using Lie algebra methods in \cite{Gibbons:2011im}.}, such as selective reflection of circularly polarised light, and it will be interesting to explore analogues of them for our new black hole solutions using linear response theory.

\section{General setup}
We consider the $D=5$ action given by
\begin{align}\label{eq:lagemcs}
S=&\int d^5 x\sqrt{-g}\left[(R+12)-\frac{1}{4} F_{\mu\nu}F^{\mu\nu}\right]-\frac{\gamma}{6}\int F\wedge F \wedge{A}\,,
\end{align}
where $F=dA$ and $\gamma$ is a constant (the boundary terms will be given later). The corresponding equations of motion are given by
\begin{align}\label{eq:eom}
&R_{\mu\nu}=-4g_{\mu\nu}+\frac{1}{2}\left(F_{\mu\rho}F_\nu{}^\rho-\frac{1}{6}g_{\mu\nu}F^2\right)\,,\nn
&d\ast F+\frac{\gamma}{2}\,F\wedge F=0\,.
\end{align}
These equations admit a unit radius $AdS_5$ vacuum solution which is dual to a class of $d=4$ CFTs with a global abelian symmetry.
The metric, $g_{\mu\nu}$, is dual to the $d=4$ energy momentum tensor, $T^{mn}$, and the gauge-field, $A_\mu$, is dual to the $d=4$ abelian current, $J^m$.
For example, 
when $\gamma=2/\sqrt{3}\approx 1.1547$ we obtain the bosonic content of $D=5$ minimal gauged supergravity, and the class
is known to include the most general class of $N=1$ SCFTs with type IIB or $D=11$ supergravity duals
\cite{Buchel:2006gb,Gauntlett:2006ai,Gauntlett:2007ma}. We will be focussing
on the range $\gamma>\gamma_c$, with
$\gamma_c\approx 1.1584$.

We will construct black hole solutions that are invariant under time translations and also have Bianchi VII$_0$ symmetry.  The Killing vectors associated with the latter are $\partial_{x_2}$, $\partial_{x_3}$, which
generate translations in the $x_2$, $x_3$ directions, respectively, and $\partial_{x_1}-k(x_2\partial_{x_3}-x_3\partial_{x_2})$,
where $k$ is a constant,
which generates a helical motion consisting of 
a translation in the $x_1$ direction combined with a simultaneous rotation in $(x_2,x_3)$ plane. The corresponding invariant one-forms are given by
\begin{align}\label{eq:one_forms}
&\omega_{1}=dx_{1},\nn
&\omega_{2}=\cos\left(kx_{1}\right)\,dx_{2}-\sin\left(kx_{1}\right)\,dx_{3},\nn
&\omega_{3}=\sin\left(kx_{1}\right)\,dx_{2}+\cos\left(kx_{1}\right)\,dx_{3},
\end{align}
which satisfy $d\omega_1=0$, $d\omega_2=-k\omega_1\wedge \omega_3$ and $d\omega_3=k\omega_1\wedge \omega_2$.
The ansatz we shall consider is given by 
\begin{align}\label{eq:ansatz}
ds^{2}&=-g\,f^{2}\,dt^{2}+\frac{dr^{2}}{g}+h^{2}\,\omega_{1}^{2}+r^{2}e^{2\alpha}\left(\omega_{2}+Qdt\right)^2+r^2e^{-2\alpha}\,\omega_{3}^{2}\,,\notag\\
A&=a\,dt+b\,\omega_2\,,
\end{align}
where $f$, $g$, $h$, $\alpha$, $Q$, $a$ and $b$ are functions of the radial coordinate $r$ only. Note that when $Q\ne 0$ 
the spacetime is stationary but not static. The black hole event horizon, located at $r=r_+$ where
$g(r_+)=Q(r_+)=a(r_+)=0$, is, generically, the non-compact Lie group Bianchi VII$_0$.

By substituting this ansatz into the equations of motion we find that $f$ and $g$ satisfy first order differential equations and that $h$, $\alpha$, $Q$, $a$ and $b$
satisfy second order equations. Furthermore, these differential equations can be obtained from substituting the ansatz directly into the action \eqref{eq:lagemcs} and then varying the seven functions of $r$. 
The constant $k$ is held fixed in these variations.
As the expressions for the equations of motion are rather long, we just record the form of the action
\begin{align}
S&=\int d^5x\,r^2hf\Bigg\{-g''-g'\left(\frac{3f'}{f}+\frac{2h'}{h}+\frac{r}{r}\right)\nn
&-\frac{2g}{r^2hf}\left[f''r^2h+f'\left(2rh+r^2h'\right)+f\left(r^2h''+2rh'+h\right)\right]\nn
&-2g(\alpha')^2-\frac{2k^2\sinh^2(2\alpha)}{h^2}+\frac{e^{2\alpha}r^2 (Q')^2}{2f^2}+\frac{k^2 r^2 e^{-2\alpha}Q^2}{2h^2f^2g}+12\nn
&+\frac{(a')^2}{2f^2}-\frac{Qb'a'}{f^2}-\frac{1}{2}\left(\frac{e^{-2\alpha}g}{r^2}-\frac{Q^2}{f^2}\right)(b')^2
-\frac{e^{2\alpha}k^2b^2}{2r^2h^2}
\Bigg\}\nn
&+\frac{\gamma k}{3}\int d^5 x\, b(ba'-ab')\,.
\end{align}
It will be useful to observe that our ansatz, and hence the equations of motion, are left invariant under the following three scaling symmetries:
\begin{align}\label{scsym}
&r\to \lambda  r,\quad (t,x_2,x_3)\to \lambda ^{-1}(t,x_2,x_3),\quad g\to \lambda ^2 g,\quad a\to \lambda a,\quad b\to \lambda b;\,\nn
&x_1\to \lambda ^{-1}x_1,\quad h\to \lambda h,\quad k\to \lambda  k\,;\nn
&t\to \lambda t,\quad f\to \lambda ^{-1} f,\quad a\to \lambda ^{-1}a,\quad Q\to \lambda ^{-1}Q\,;
\end{align}
where $\lambda $ is a constant.

The equations of motion admit the electrically charged AdS-Reissner-N\"ordstrom black brane solution. It has $h=r$, $f=1$, $\alpha=Q=b=0$, and hence,
\begin{align}\label{eq:ansatzrn}
ds^{2}=-gdt^{2}+\frac{dr^{2}}{g}+r^{2}\left(dx_1^2+dx_2^2+dx_3^2\right),\qquad
A=a\,dt\,,
\end{align}
with
\begin{align}\label{gaval}
g=r^{2}-\frac{r^{4}_{+}}{r^{2}}+\frac{\mu^{2}}{3}\,\left(\frac{r_{+}^{4}}{r^{4}}-\frac{r^{2}_{+}}{r^{2}} \right),\qquad a=\mu\,\left(1-\frac{r_{+}^{2}}{r^{2}} \right)\,.
\end{align}
The AdS-RN black brane is static and has Euclidean, $ISO(3)$, symmetry. It 
has temperature $T=(6r_+^2-\mu^2)/6\pi r_+$ and
describes the high temperature, spatially homogeneous and isotropic phase of the dual CFTs when held at finite chemical potential $\mu$ with respect to the 
global abelian symmetry.

\subsection{Asymptotic $AdS_{5}$ and near-horizon expansions}
We will be interested in new black hole solutions that asymptotically approach $AdS_5$ in the UV and are dual to $d=4$ phases 
where the breaking of the Euclidean symmetry to a helical order is spontaneously generated.
By analysing the equations of motion we can construct the following asymptotic expansion as $r\to\infty$:
\begin{align}\label{asexp}
&g=r^{2}\,\left(1-\frac{M}{r^{4}}+\cdots \right)\notag\\
&f=f_{0}\left(1-\frac{c_{h}}{r^{4}}+\cdots\notag\right)\\
&h=r\,\left(1+\frac{c_{h}}{r^{4}}+\cdots \right)\notag\\
&\alpha=\frac{c_{\alpha}}{r^{4}}+\cdots\notag\\
&Q=f_0\left(\frac{c_{Q}}{r^{4}}+\cdots\right)\notag\\
&a=f_{0}\,\left(\mu+\frac{q}{r^{2}}+\cdots\right)\notag\\
&b=\frac{c_{b}}{r^2}+\cdots
\end{align}
At a convenient juncture we will use the scaling symmetries \eqref{scsym} to set $f_0=\mu=1$.
The UV data is then specified by seven parameters $M,c_h,c_\alpha,c_Q,q,c_b$ and $k$. Note that
we have fixed the asymptotic fall-off of $h$ in \eqref{asexp} so we can no longer use \eqref{scsym} 
to scale $k$. The holographic
interpretation of these parameters will be discussed later.

At the black hole horizon, located at $r=r_{+}$,  the functions have the analytic expansion
\begin{align}\label{nhexp}
g&=g_{+}\,\left(r-r_{+}\right)+\dots\notag\\
f&=f_{+}+\dots\notag\\
h&=h_{+}+\dots\notag\\
\alpha&=\alpha_{+}+\dots\notag\\
Q&=Q_{+}(r-r_+)+\dots\notag\\
a&=a_{+}\,\left(r-r_{+}\right)+\dots\notag\\
b&=b_{+}+\dots
\end{align}
Regularity of the metric at the black hole horizon can easily be seen by using
the in-going Eddington-Finkelstein coordinates $v,r$ where $v\approx t+(g_+f_+)^{-1}\ln(r-r_+)$.
The full expansion is fixed in terms of the seven constants $f_+$, $\alpha_{+}$, $h_{+}$, $Q_{+}$, $a_{+}$, $b_+$ and $r_+$.
In particular, the coefficient $g_+$ is fixed by these constants:
\begin{align}
g_+=-\frac{e^{2 \alpha_+} k^2 b_+^2}{12h_+^2 r_+}
+\left(4-\frac{a_+^2}{6 f_+^2}\right) r_+
-\frac{e^{2 \alpha_+} Q_+^2 r_+^3}{4 f_+^2}\,.
\end{align}

After fixing the scaling symmetries \eqref{scsym}
we have seven UV parameters and seven IR parameters. We have two first order differential equations and five which are second order, so
a solution is fixed by twelve parameters. Thus, generically, we
expect a two parameter family of black hole solutions, which we will label by $k$ and temperature $T$.

\subsection{Thermodynamics}\label{thermo}
To analyse the thermodynamics of the black hole solutions we will need to calculate the on-shell Euclidean action. 
Additional details are presented in appendix \ref{moreonthermo}.
We
analytically continue by setting $t=-i\tau$ and in order to get a real metric and vector field we should also
write $Q_+=i\bar Q_+$, $a_+=i\bar a_+$. Near $r=r_+$ the Euclidean solution
then takes the approximate form
\begin{align}\label{eq:eulcid}
ds^{2}_E&\approx g_+\,f_+^{2}(r-r_+)\left(d\tau+\frac{\bar Q_+r_+^2e^{2\alpha_+}}{g_+f_+^2}\omega_2\right)^2
+\frac{dr^{2}}{g_+(r-r_+)}\nn
&\qquad\qquad\qquad\qquad\qquad\qquad\qquad+h_+^{2}\,\omega_{1}^{2}+r_+^{2}e^{2\alpha_+}\left(\omega_{2}\right)^2+r_+^2e^{-2\alpha_+}\,\omega_{3}^{2}\,,\notag\\
A&\approx\bar a_+(r-r_+)\,d\tau+b\,\omega_2\,.
\end{align}
Regularity of the solution at $r=r_+$ is then easily seen by making the coordinate change $\rho=2g_+^{-1/2}(r-r_+)^{1/2}$ and making $\tau$ periodic with period
$\Delta\tau=4\pi/(g_+f_+)$, corresponding to temperature $T=\left(f_{0}\Delta\tau\right)^{-1}$. We can also read off the area of the event horizon and,
since we are working in units with $16\pi G=1$, we deduce that the entropy density is given by
\begin{align}
s=4\pi r^{2}h_{+}\,.
\end{align}

We will consider the total Euclidean action, $I_{Tot}$, defined as
\begin{align}
I_{Tot}=I+I_{bndy}\,,
\end{align}
where $I=-iS$ and $I_{bndy}$ is the Euclidean boundary action of \cite{D'Hoker:2009bc}, including counter-terms,
which is given explicitly in appendix \ref{moreonthermo}.
We next define the potential $W$, and a corresponding density $w$, for the grand canonical ensemble via $W=T[I_{Tot}]_{OS}=w\mathrm{vol}_3$, where $[I_{Tot}]_{OS}$ is the on-shell Euclidean action and $\mathrm{vol}_3=\int dx^1 dx^2 dx^3$.
A calculation reveals that $w$ can be expressed in two equivalent ways
\begin{align}\label{osact}
w&=-M =3M+8c_{h}+2\mu q-Ts\,,
\end{align}
with the equality of the two expressions giving a Smarr type formula.

A variation of the bulk action $I$ gives equations of motion and boundary terms. 
Thus an on-shell variation only gets contributions from the boundary. We hold $k$ fixed in these
variations, for reasons we discuss in the next subsection, and for the Euclidean black hole we then only get contributions at $r\to\infty$. 
Combining this with an on-shell variation of the boundary action
$I_{bndy}$ and using the asymptotic expansion \eqref{asexp} we deduce that $w=w(T,\mu)$ and
\begin{align}\label{delw}
\delta w= -s\delta T +2 q\delta\mu\,.
\end{align}

To illuminate the holographic meaning of the constants appearing in the UV expansion \eqref{asexp}, 
we now compute the expectation value of boundary stress-energy tensor and the current. The relevant terms for the stress tensor are given by\cite{Balasubramanian:1999re}
\begin{align}
\langle T_{mn}\rangle=\lim_{r\to\infty}r^2\left[-2K_{mn}+2(K-3)(g_\infty)_{mn}+\dots\right]\,.
\end{align}
Using \eqref{asexp} we obtain
\begin{align}\label{emtvev}
\langle T_{tt}\rangle&=\,f_{0}^{2}\,\left(3M+8c_{h}\right)\notag\\
\langle T_{tx_{2}}\rangle&=4f_0c_{Q}\,\cos\left(k x_{1}\right)\notag\\
\langle T_{tx_{3}}\rangle&=-4f_0c_{Q}\,\sin\left(k x_{1}\right)\notag\\
\langle T_{x_{1}x_{1}}\rangle&=\,\left(M+8c_{h}\right)\notag\\
\langle T_{x_{2}x_{2}}\rangle&=\,\left(M +8c_{\alpha}\,\cos\left(2kx_{1} \right) \right)\notag\\
\langle T_{x_{3}x_{3}}\rangle&=\,\left(M -8c_{\alpha}\,\cos\left(2kx_{1} \right) \right)\notag\\
\langle T_{x_{2}x_{3}}\rangle&=-8\,c_{\alpha}\,\sin\left(2kx_{1}\right)\,.
\end{align}
One can check that this is traceless $(g_\infty)^{mn}\langle T_{mn}\rangle=0$.
Setting $f_0=1$ we see that the energy density of our solutions, $\varepsilon$, is given by
\begin{align}
\varepsilon=3M+8c_{h}\,.
\end{align}
Furthermore, we also deduce from \eqref{osact} and \eqref{delw} that the first law can also be written in the form:
\begin{align}\label{plop}
\delta\varepsilon=T\delta s-2\mu\delta q\,.
\end{align}
Returning to \eqref{emtvev} we see that $c_\alpha$ specifies spatially modulated pressures and shear in  
the $(x_2,x_3)$ plane, with the length scale of the modulation fixed by the wave-number $k$. The pressure in the $x_1$ direction is given by $M+8 c_h$. If we define $\bar p$ to be the average of the three pressures we have
$\bar p=M+8/3 c_h$, and the Smarr formulae in
\eqref{osact} can be written $\varepsilon+2\mu q=Ts+\bar p-8/3 c_h$. These features were also seen for the black holes found
in \cite{Donos:2012gg}. A new feature of the black holes we are considering here is that there is spatially modulated momentum in the $(x_2,x_3)$ plane specified by
$c_Q$ and $k$.

We next calculate the expectation value of the current. The relevant terms are given by \cite{D'Hoker:2009bc}
\begin{align}
\langle J_{m}\rangle=\lim_{r\to\infty}r^3\left[F_{rm}+\dots\right]\,.
\end{align}
where the ellipsis refer to terms that will not be relevant here.
Using \eqref{asexp} we obtain
\begin{align}\label{vevofJ}
\langle J_{t}\rangle&=-2f_0q\notag\\
\langle J_{x_1}\rangle&=0\notag\\
\langle J_{x_2}\rangle&=-2c_b\cos(kx_1)\nn
\langle J_{x_3}\rangle&=2c_b\sin(kx_1)\,.
\end{align}
From the temporal component we see that the constant $q$ fixes the charge density. From the spatial components we see that $c_b$ fixes the strength of the spontaneously generated spatially modulated helical current. It is clearly circularly polarised.

\subsection{Variations of $k$}
In the variations to get \eqref{delw}, or equivalently \eqref{plop}, we held the wave-number $k$ fixed.
One can consider arbitrary non-normalisable and normalisable deformations of the fields and then
expand them in a complete basis of functions. Here we are viewing $k$ as labelling one of the modes and
hence should not be varied to obtain the equations of motion\footnote{An analogous 
procedure was employed in \cite{Donos:2012gg} and also, essentially in a field theory context, in \cite{Ooguri:2010kt}. To clarify this point, in appendix \ref{moreonthermo} we consider a more general set-up in which
we also allow a more general deformation parameter at infinity. Specifically, we consider 
$b=\mu_{b}+\frac{c_{b}}{r^{2}}+\ldots$, with
$\mu_b$ and $c_b$ being the non-normalisable and normalisable deformations of the magnetic part of the 
gauge field for a particular mode $k$.}. Furthermore, it should not be varied to obtain an on-shell variation of the action in order to obtain the thermodynamics for a particular solution labelled by $k$.
As we discuss further in the next section, we will obtain a two parameter family of black hole solutions to the equations of motion that depend on $k$ and $T$ (see figure \ref{fig:1}). At fixed temperature, this should be viewed as a moduli space of
solutions, labelled by $k$ and we should choose the solution labelled by a specific value of $k$ that has the smallest free energy $w$. This leads to a one-parameter family of thermodynamically preferred solutions labelled
by $T$, given by the red line in figure \ref{fig:1}. In fact this red line is specified by the condition that the action $I_{Tot}$ is stationary with respect to a free variation of $k$:
\begin{align}\label{eq:deltak}
\int_{r_+}^{\infty} dr\Bigg\{
k\left(\frac{4r^2 f\sinh^2(2\alpha)}{h}
-\frac{r^4 e^{-2\alpha}Q^2}{fgh}+\frac{e^{2\alpha}b^2f}{h}\right)
-\frac{\gamma }{3} b(ba'-ab')\Bigg\}=0\,.
\end{align}
In appendix \ref{moreonthermo} we will discuss how this arises from contributions to varying the action at $x_1=\pm \infty$. 

Another perspective is to consider the $x_1$ direction to be periodic with $x_1\cong x_1+L$. In this case we should only consider discrete wave-numbers $k=n/(2\pi L)$, for arbitrary integer $n$, and there
is no issue of varying $k$ to obtain the equations of motion. 
In this case there will be a discrete set of solutions on figure \ref{fig:1} and, at a fixed
temperature, one should just choose the one with smallest free energy, as usual. One finds that the system
will, in general, jump discontinuously from one branch to another giving a series of first order phase transitions.
In the limit that $L\to\infty$ we will recover the continuum picture that we have discussed above.

It is worth noting that the same kinds of issues also arise for homogeneous and isotropic phases.
For example, recall the basic $s$-wave holographic superconducting black holes \cite{Gubser:2008px,Hartnoll:2008kx}. In this setting 
the AdS-RN black brane, which describes
the high temperature phase, becomes unstable to the formation of charged scalar hair. Although only scalar modes with $k=0$ have
been discussed in the literature, a linearised analysis for $k\ne 0$ will produce a curve analogous to that in figure \ref{fig:lin} but it will now be symmetric about $k=0$.
Hence, there should also be a two-parameter family of superconducting black hole solutions labelled by $T$ and $k$. In this case, however, the thermodynamically
preferred curve of solutions will just be the solution with $k=0$.

\section{Helical black holes}
\begin{figure}[t!]
\centering
\includegraphics[width=0.4\textwidth]{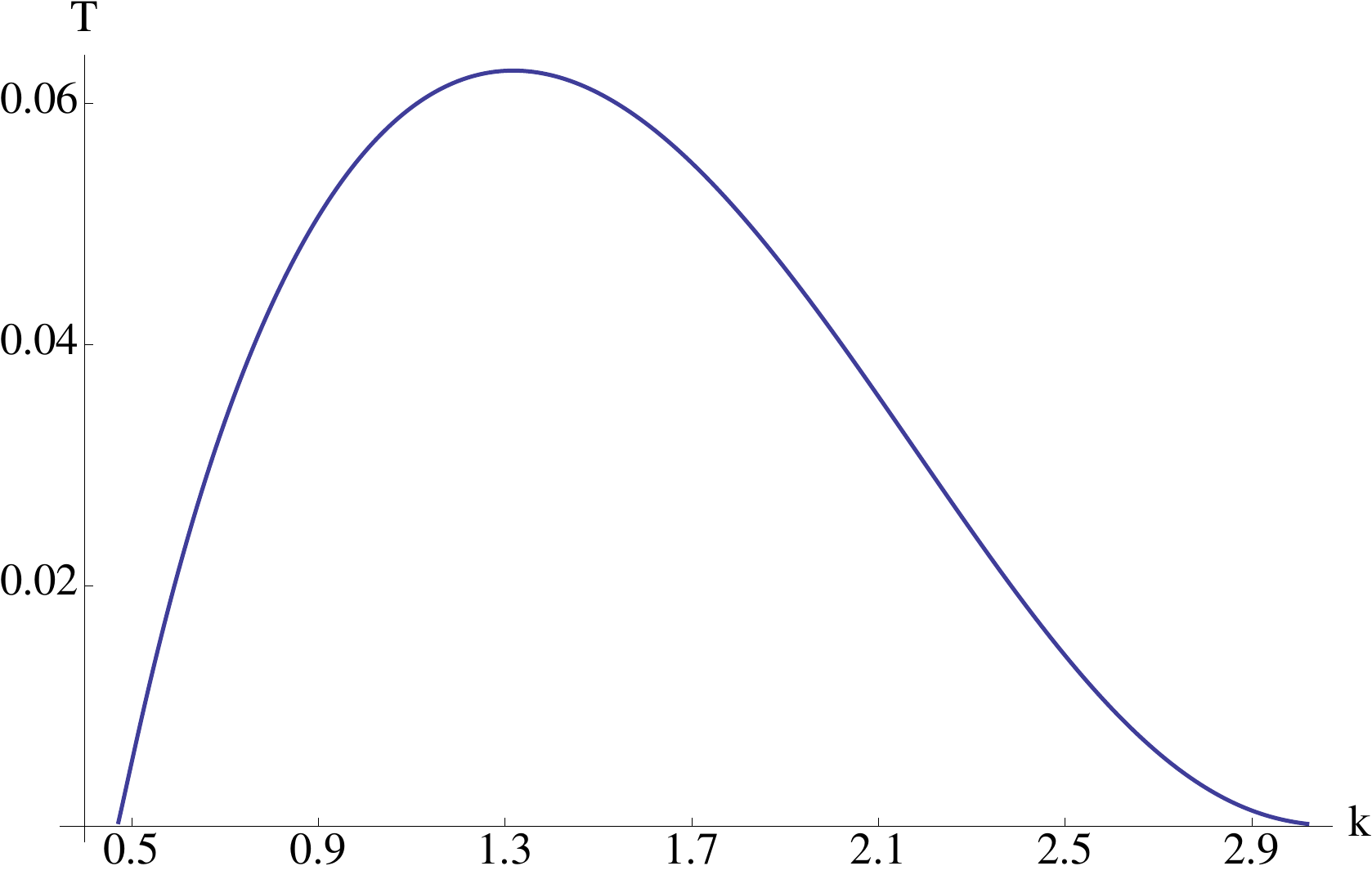}
\caption{The curve denotes the critical temperature at which the AdS-RN black brane becomes unstable and also
where the new branches of helical black holes, given in figure \ref{fig:1}, appear. The plot is for $\gamma=1.7$ and $\mu=1$.
\label{fig:lin}}
\end{figure}

Based on the analysis of linearised perturbations about the AdS-RN black brane solution carried out in 
\cite{Nakamura:2009tf} we expect to be able to construct spatially modulated black hole solutions provided that the Chern-Simons
coupling $\gamma$ is larger than $\gamma_c\approx 1.1584$. 
We will now set $\gamma=1.7$, and hence $\gamma/\gamma_c\approx 1.47$, but we have checked that several other values lead to
qualitatively similar results. For this value the linearised analysis of \cite{Nakamura:2009tf}, which we summarise in appendix \ref{linal},
leads to the curve
presented in figure \ref{fig:lin} which denotes, for a given value of $k$, the temperature at which the
AdS-RN black brane becomes unstable. Hence for $k$ in the range $0.47 \lesssim k\lesssim3.05$ we expect
to be able to find the new black hole solutions.

\begin{figure}[t!]
\centering
\includegraphics[width=0.5\textwidth]{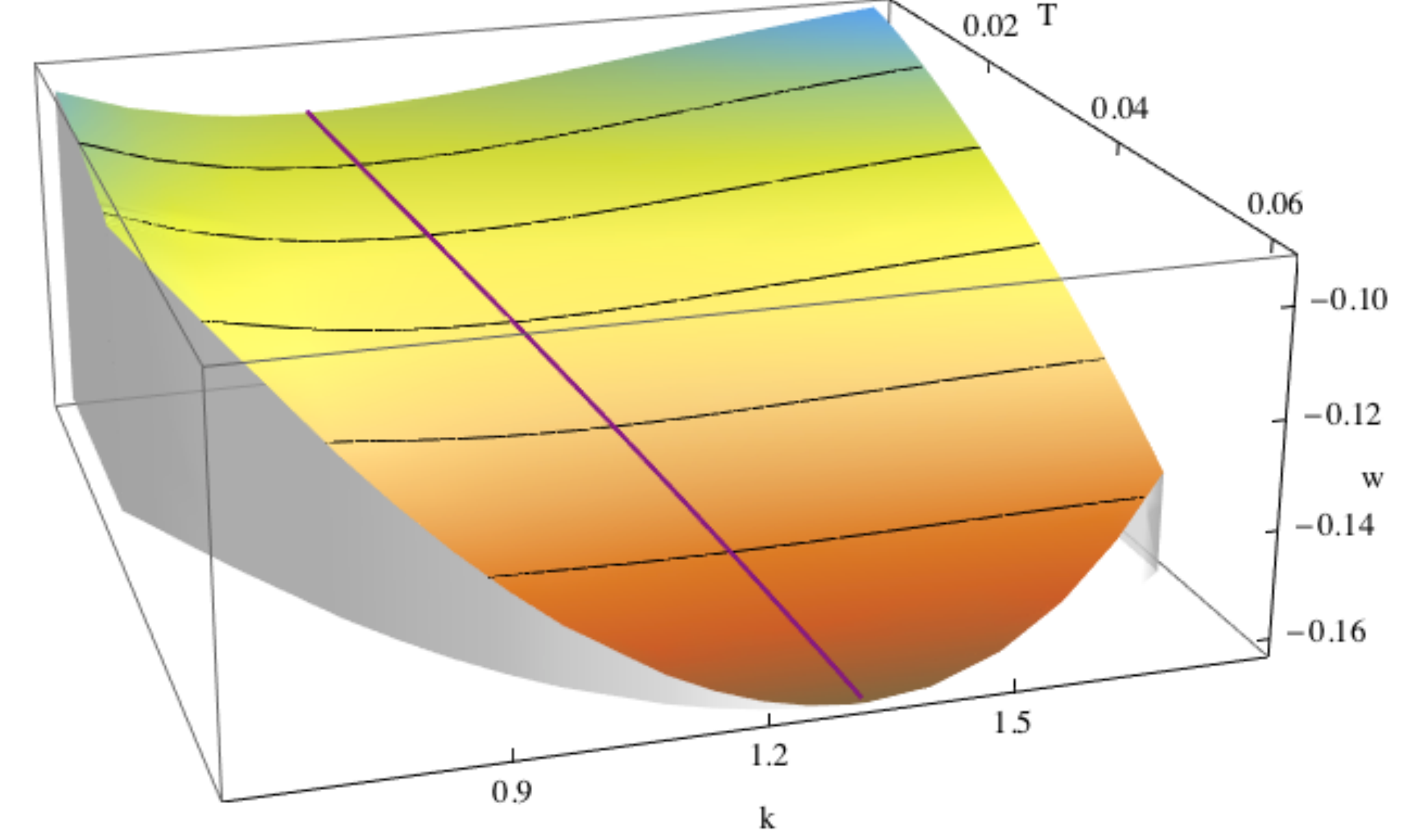}
\caption{The two-parameter family of helical black holes, labelled by temperature $T$ and wave-number $k$, and their free-energy $w$.
The red line denotes the thermodynamically preferred locus, which minimises $w$ over the moduli space of solutions at fixed $T$ labelled by $k$. 
The plot is for $\gamma=1.7$ and $\mu=1$.
\label{fig:1}}
\end{figure}

\begin{figure}[t!]
\centering
\includegraphics[width=0.4\textwidth]{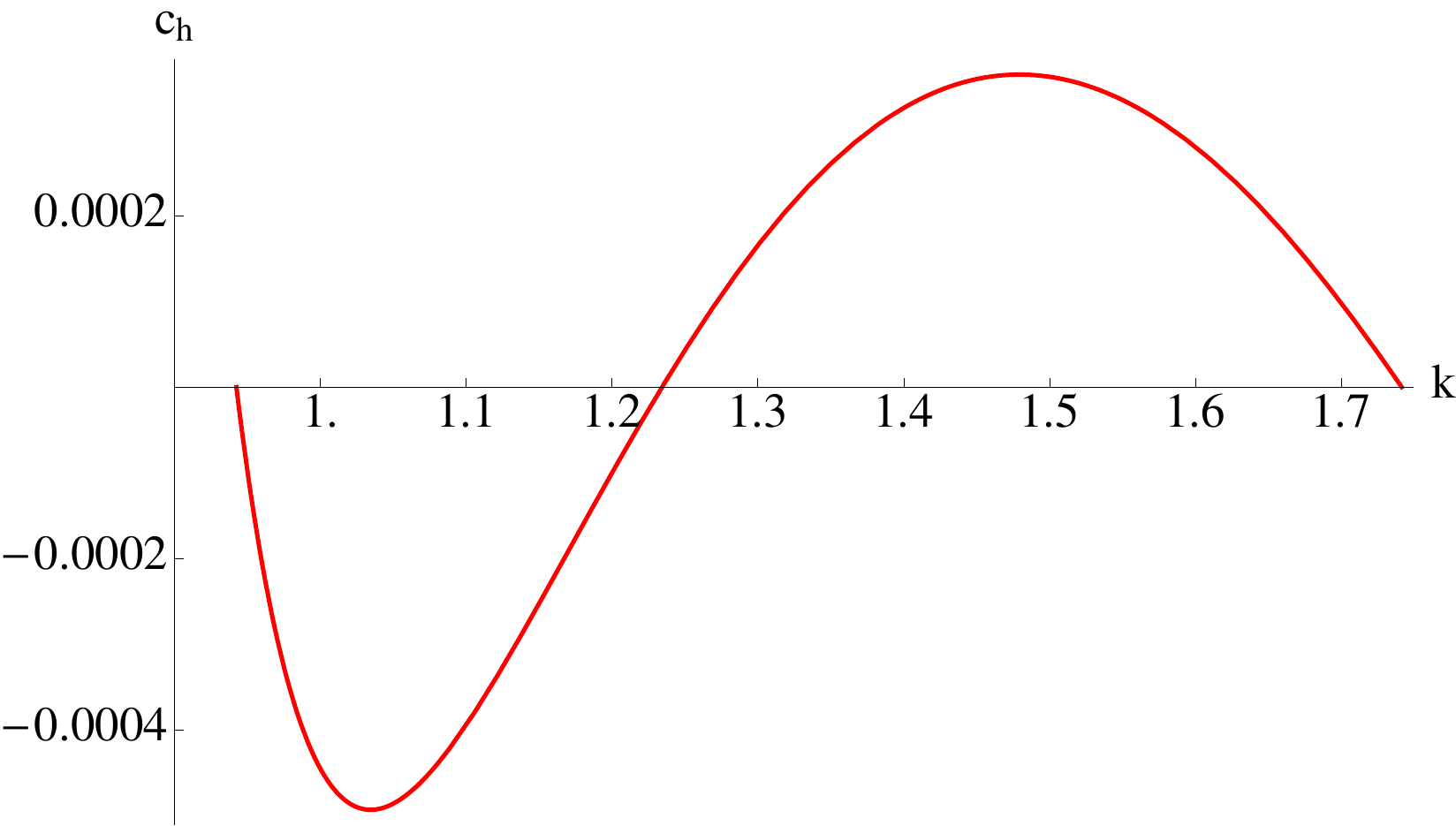}
\caption{Plot of $c_h$ versus $k$ for the one parameter family of helical black hole solutions given in
figure \ref{fig:1} for the representative temperature $T\approx 0.0535$. In particular $c_h=0$ on the thermodynamically
preferred red line of solutions given in \ref{fig:1}. The plot is for $\gamma=1.7$ and $\mu=1$.
\label{fig:ch}}
\end{figure}

\begin{figure}[t!]
\centering
\includegraphics[width=0.4\textwidth]{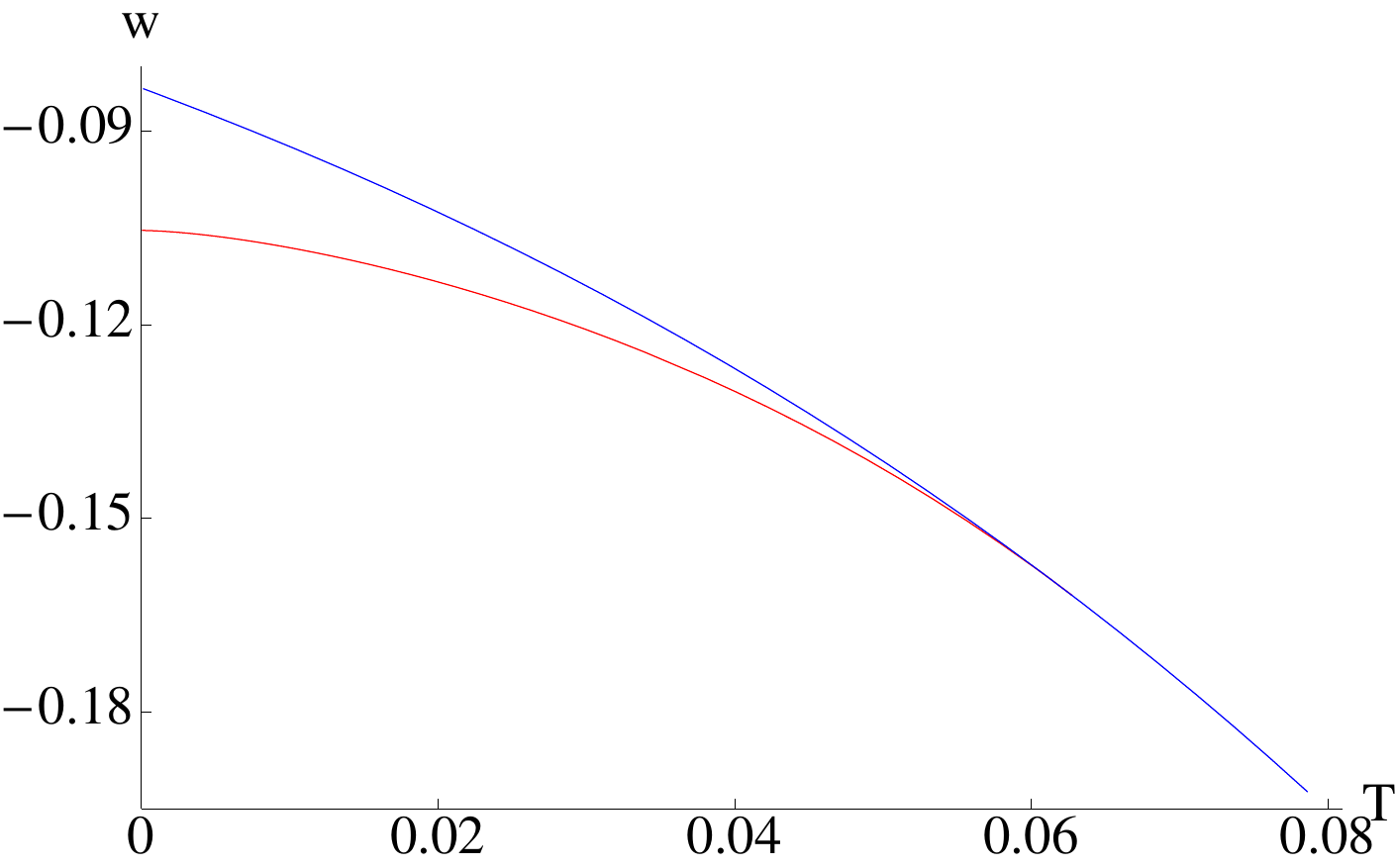}
\hskip1em \includegraphics[width=0.4\textwidth]{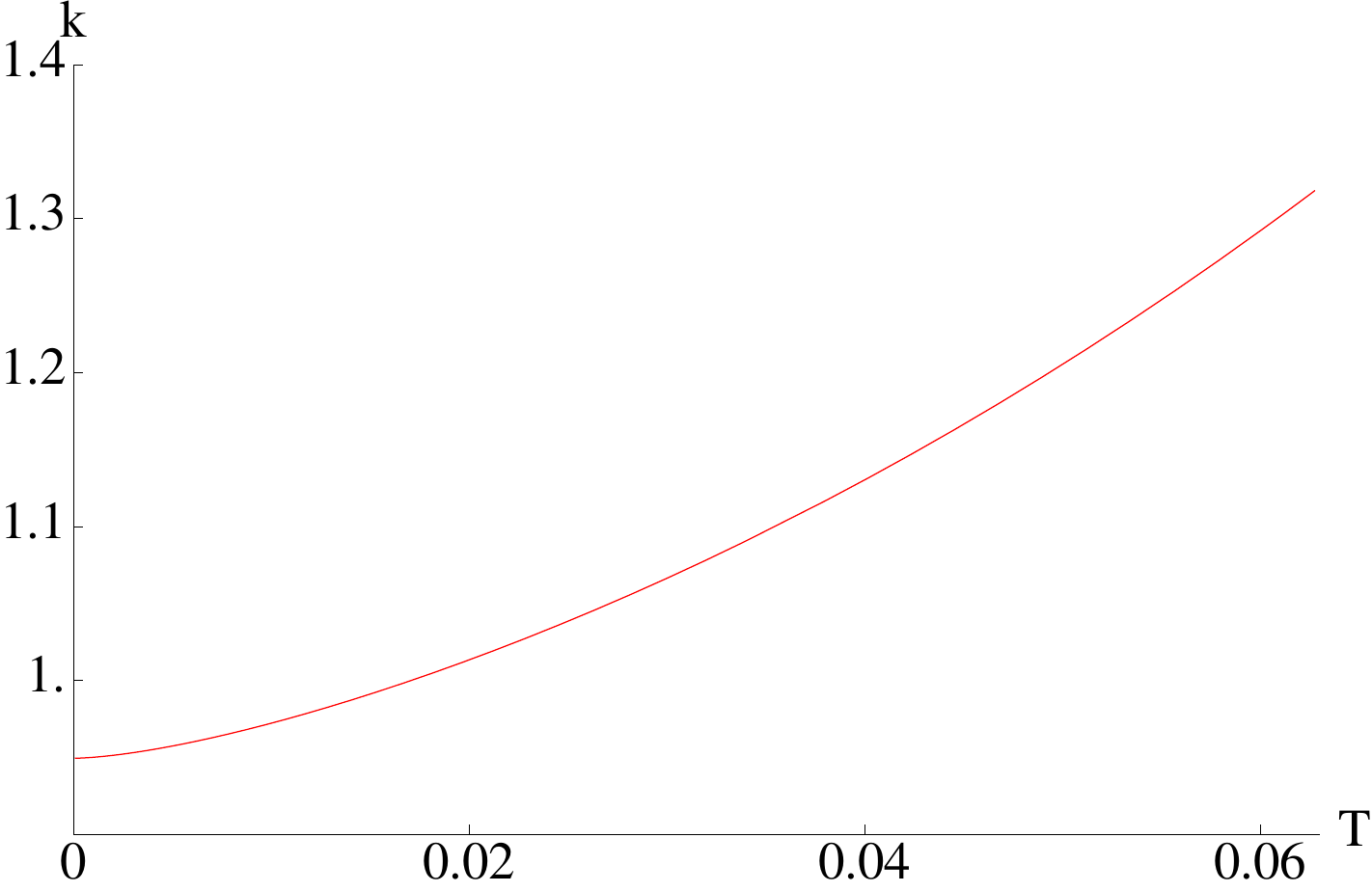}
\includegraphics[width=0.4\textwidth]{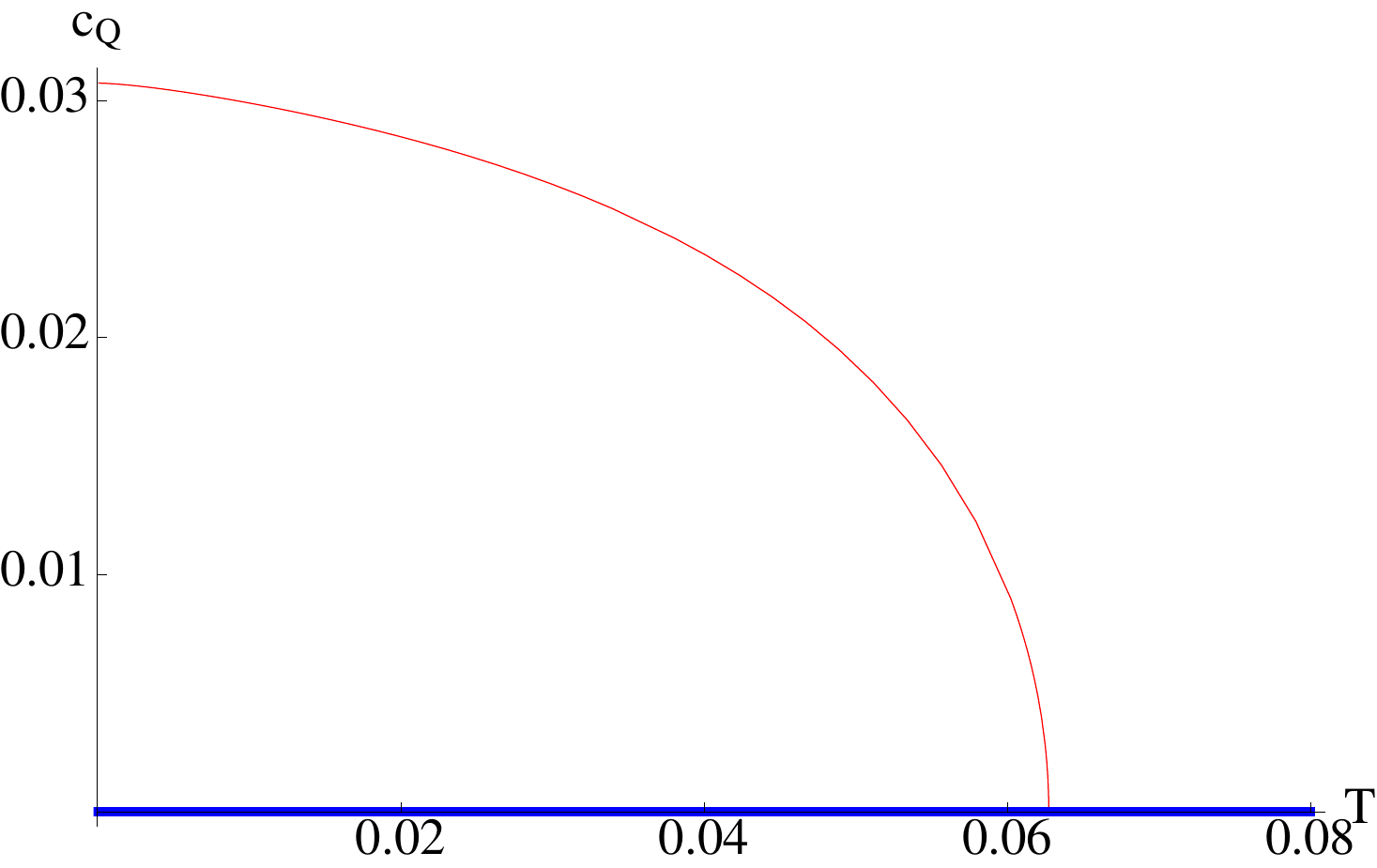}
\hskip1em \includegraphics[width=0.4\textwidth]{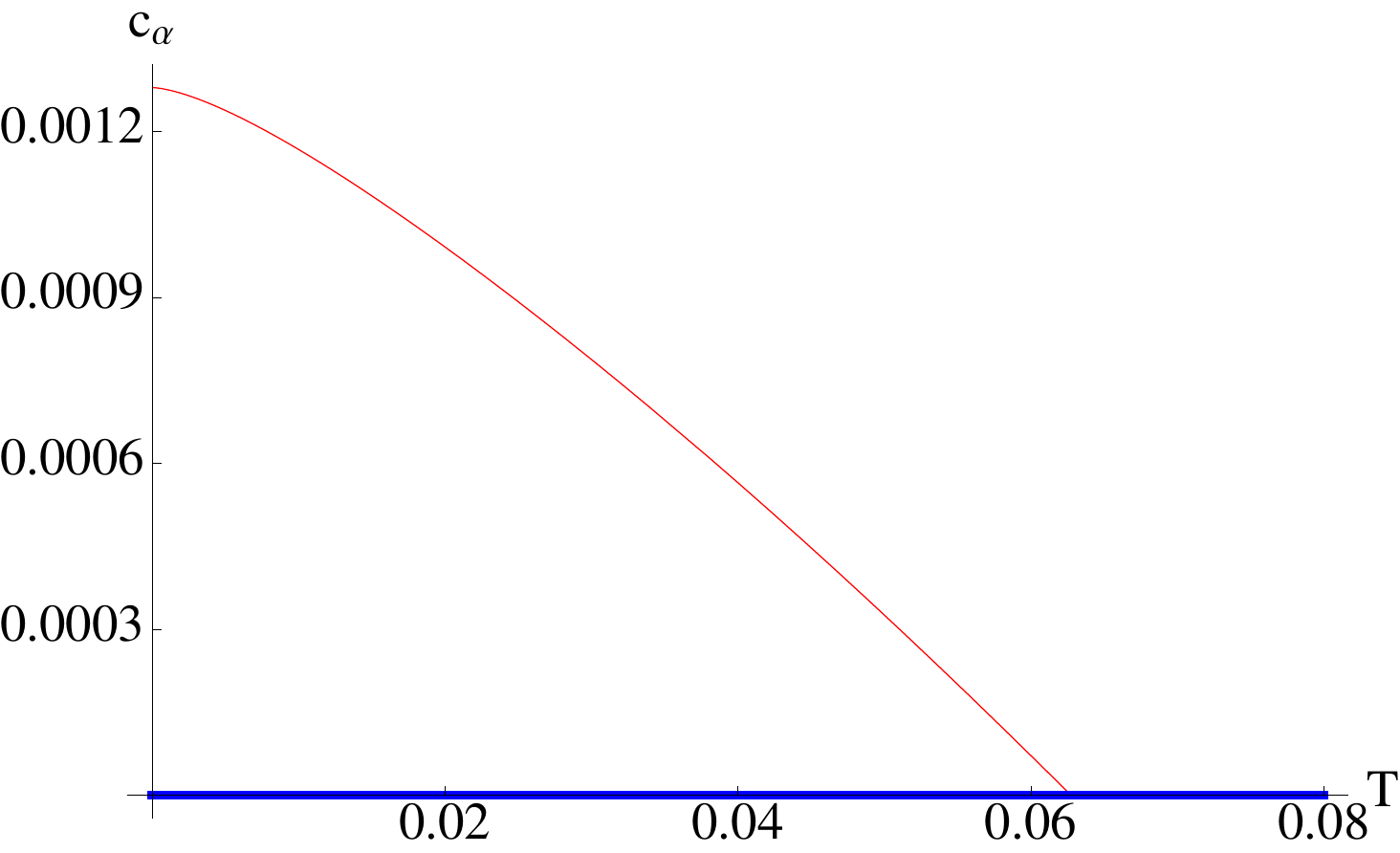}
\includegraphics[width=0.4\textwidth]{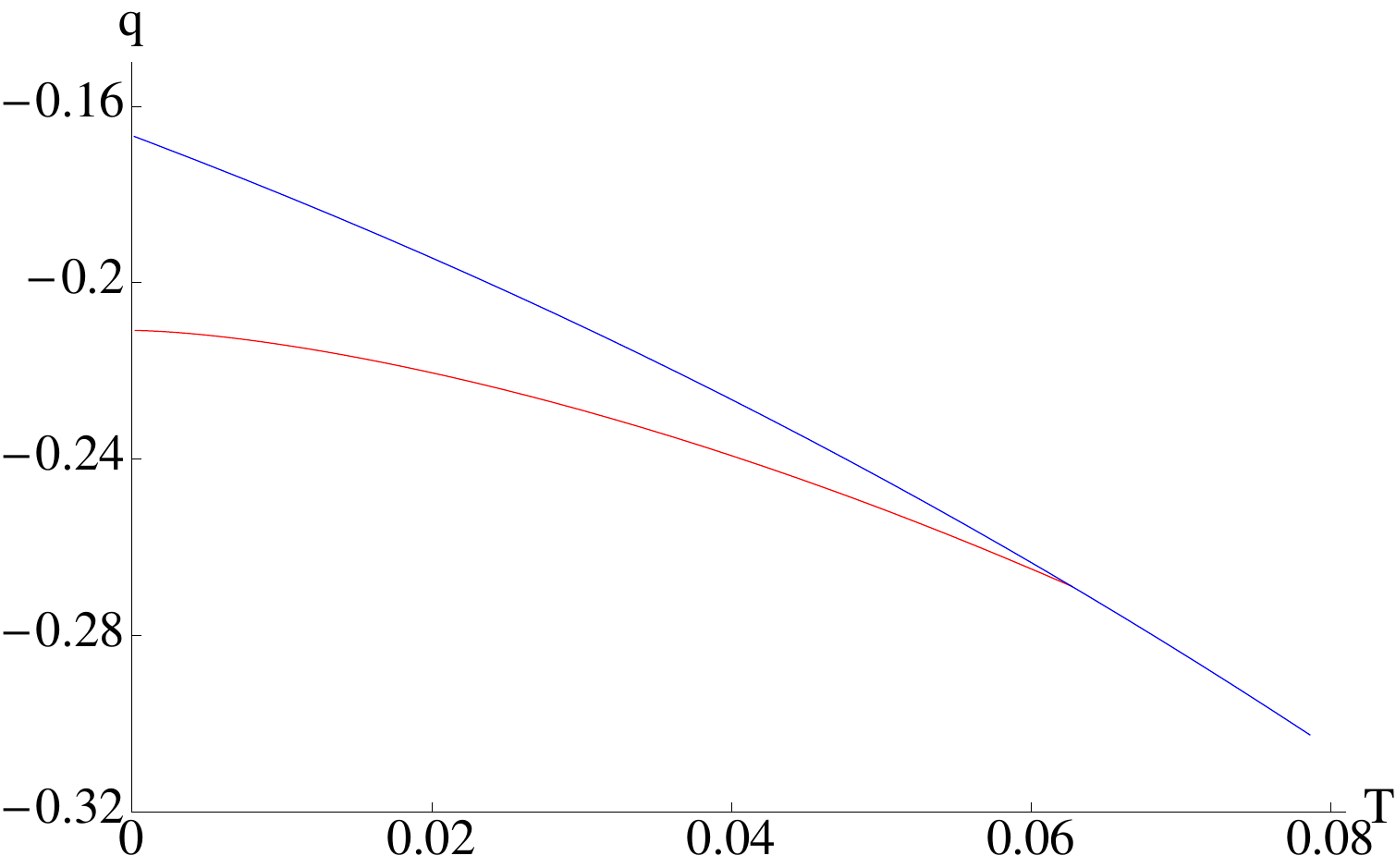}
\hskip1em \includegraphics[width=0.4\textwidth]{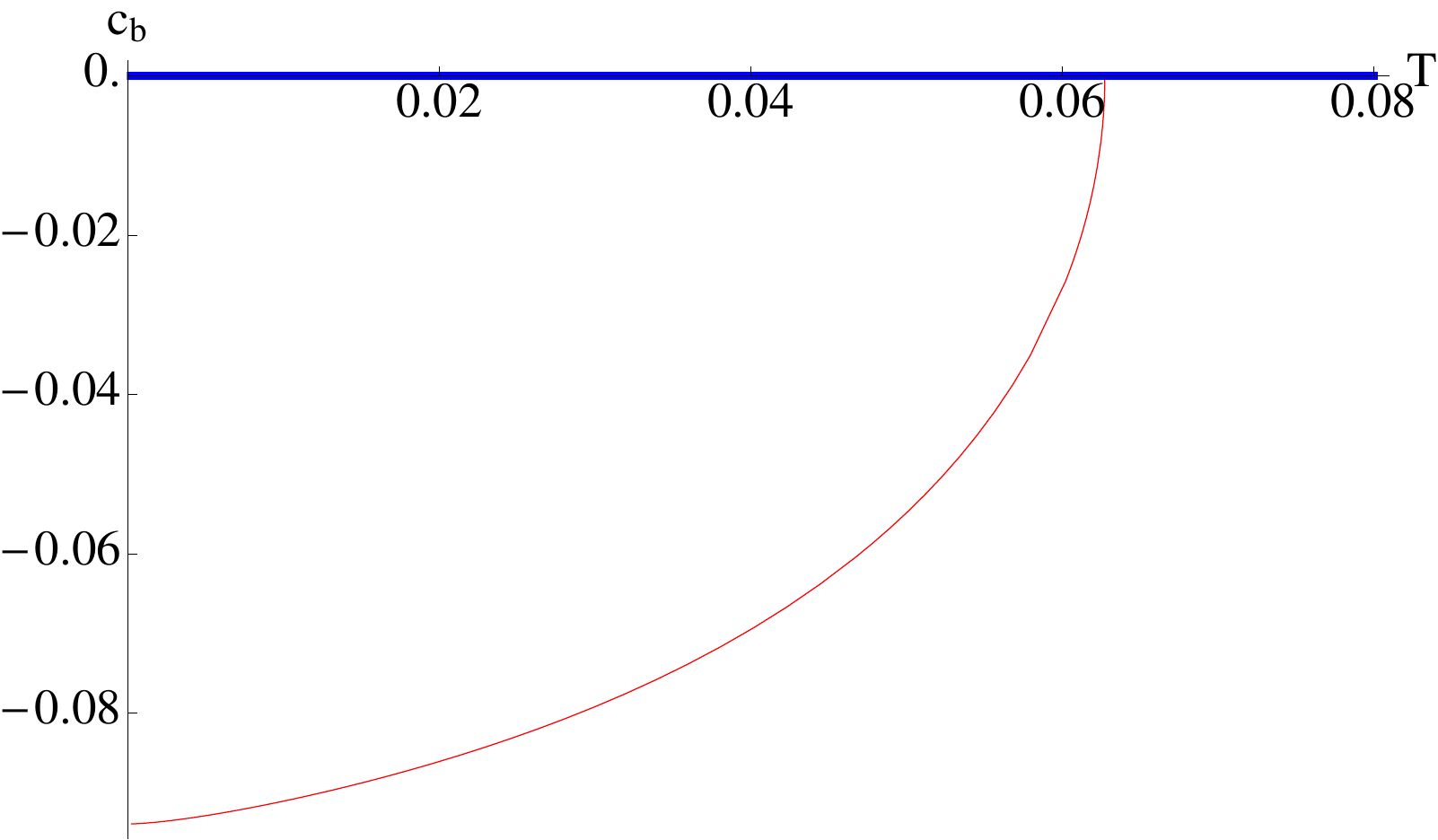}
\includegraphics[width=0.4\textwidth]{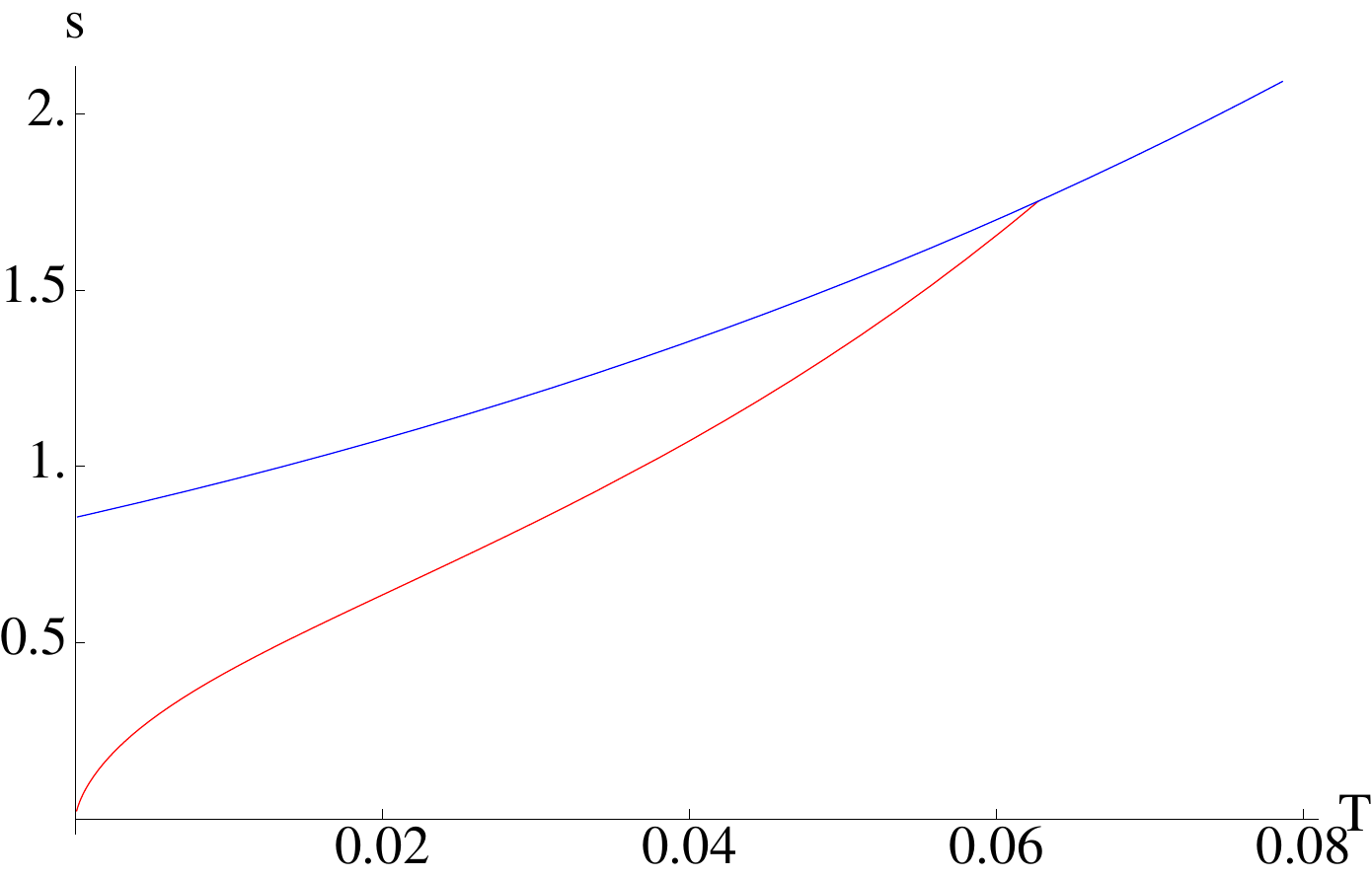}
\caption{The red lines plot various physical quantities against temperature $T$ for the thermodynamically preferred helical black hole solutions on the red line in figure \ref{fig:1}. The blue line refers to the AdS-RN black hole solution. $w$ is the free energy and $k$ is the wave-number of the helical order.
$c_Q$ and $c_\alpha$ fix the spatially modulated momentum and stress/strain in the $(x_2,x_3)$ plane, respectively.
$q$ and $c_b$ determine the size of the charge and the spatially modulated current, respectively, 
and $s$ is the entropy density. The plots are for $\gamma=1.7$ and $\mu=1$.
\label{fig:2}}
\end{figure}

The new helical black hole solutions are obtained by solving the equations of motion numerically for the ansatz \eqref{eq:ansatz} 
with boundary conditions at the asymptotic $AdS_5$ boundary given in \eqref{asexp}, 
and at the black hole horizon given in \eqref{nhexp}. We use the scaling symmetries \eqref{scsym} to set $f_0=\mu=1$. As mentioned earlier a simple
parameter count indicates that we expect, generically, a two-parameter family of solutions which we take to be labelled by temperature $T$ and wave-number $k$. In practise we
fix a specific value of $k$ and then construct a one parameter family of solutions labelled by the temperature $T$. We considered twenty different values of $k$, in the range $0.6 \leq k\leq 1.8$ (focussing on the peak of the curve in figure \ref{fig:lin}),
and we have displayed our results in figures 
\ref{fig:1} - \ref{fig:2}.

Figure \ref{fig:1} shows the two-parameter family of solutions and their free energy $w$. We first note that the boundary of the surface projected onto the $(k,T)$ plane reproduces the curve of critical temperatures as a function of $k$ where the AdS-RN black brane becomes unstable given in figure \ref{fig:lin}. 
We next note that for any fixed temperature the helical black holes have less free energy than the AdS-RN black hole for any value of $k$. Thus, from figure \ref{fig:1} we deduce that there is a second order phase transition at $T=T_c\approx 0.0627$ at $k=k_c\approx 1.32$ with the system moving from a homogeneous and isotropic phase to a spatially modulated helical phase. As the temperature is lowered we need to find the value of $k$ for which the black hole has the lowest free energy.
This leads to the one-parameter family of thermodynamically preferred black hole solutions which are marked with
a red line in figure 1. 
We have also checked numerically that this red line coincides with imposing \eqref{eq:deltak}.
Interestingly, our numerical analysis indicates that all of the black hole solutions on the red line
have $c_h=0$ (a similar phenomenon was also seen in \cite{Donos:2012gg}). In figure \ref{fig:ch} we have plotted the behaviour of
$c_h$ versus $k$ for a representative temperature of $T\approx 0.535$ and we see that it vanishes along the red curve as well as on 
on the boundary curve of figure \ref{fig:1}. It would be interesting to understand the underlying reason for this behaviour.

In figure \ref{fig:2} we show the behaviour of various other physical quantities as a function of temperature for the
thermodynamically preferred branch, given by the red line in figure \ref{fig:1}, and marked with red lines in figure \ref{fig:2}, as well as for the AdS-RN black hole solution, marked with blue lines in figure \ref{fig:2}. The first two panels show the free energy $w$ for both solutions and the wave-number $k$ for the red line.
The pitch, $p$ of the helical order is given by $p=2\pi/k$ and hence we see that the 
pitch monotonically increases as the temperature is decreased.
The next two panels show the behaviour of $c_Q$ and $c_\alpha$ which, we recall from \eqref{emtvev}, determine the strength of the spatially modulated momentum and stress/strain in the  $(x_2,x_3)$ plane, respectively.
The next two panels show the behaviour of the charge density $q$ and also the behaviour of $c_b$, which we recall from \eqref{vevofJ}, determines the strength of the spatially modulated helical current. The final panel shows the behaviour of the entropy density, $s$.
By analysing the behaviour close to the critical temperature, $T=T_c\approx 0.0627$, we find the following mean-field behaviour
\begin{align}
& c_Q\approx 2953\, T_{c}^{4}\,\left(1-\frac{T}{T_{c}} \right)^{1/2}, \quad c_\alpha \approx 107\, T_{c}^{4} \left(1-\frac{T}{T_{c}} \right),\nn
&c_b\approx 530\, T_{c}^3\,\left(1-\frac{T}{T_{c}} \right)^{1/2}\qquad
  k\approx T_{c}\,\left(11.5+9.58\,\frac{T}{T_c} \right).
\end{align}

\subsection{Low temperature behaviour}
It is clear from figures \ref{fig:1} and \ref{fig:2} that the helical order persists as $T\to 0$, and in particular $k$ approaches
a non-zero value in this limit. Furthermore, the entropy density approaches zero. 
However, we have not yet been able to pin down the precise behaviour of the solution in this limit, and we leave this interesting issue for further instigation. However, we record a couple of conclusions based on our numerical results. Amongst the coefficients in the near horizon expansion \eqref{nhexp}, we find that $h_{+}\to\infty$ and $\alpha_{+}\to -\infty$,
with the entropy density, $s=4\pi r_{+}^{2}h_{+}$, going to zero and 
$r_{+}h_{+}e^{-\alpha_{+}}$ approaching a constant value. We also find that $f_+,a_+$ vanish while $Q_+, b_+$ go to constant values.
Starting from the phase transition the value of $F_{\mu\nu}F^{\mu\nu}$ at the horizon is $-8/r_+^2$ and this monotonically increases and approaches 24 as $T\to 0$. Thus associated with the vanishing entropy density, we have an expulsion of electric charge somewhat
reminiscent of \cite{D'Hoker:2012ej}.
We also evaluated the horizon value of the Ricci scalar and the square of the Ricci tensor and it appears that $R\to -18$ and $R_{\mu\nu}R^{\mu\nu}\to 108$. 
It is tantalising that that these are the same values for the Schr\"odinger solution of \cite{D'Hoker:2010ij}.

\section{Final Comments}
We have constructed, numerically, a new class of electrically charged AdS$_5$ black holes that are dual to $d=4$ CFTs at finite charge density
acquiring a helical current order via a second order phase transition. We have extracted a number of physical properties of the
helical phase including the temperature dependence of the wave-number $k$ that fixes the pitch, $p=2\pi/k$, of the helix. We have shown
that the pitch monotonically increases as the temperature is lowered but approaches a finite value as $T\to 0$. 
Furthermore, our numerical results indicate that the entropy density goes to zero in this limit. It will be very interesting to further analyse
the precise behaviour of our solutions as $T\to 0$ in order to better understand the emergent spatially modulated ground state at $T=0$.

Another direction is to calculate various transport coefficients by calculating various two point functions. There will be a number of different channels
to analyse and we expect a rich structure. It will also be interesting to explore the related hydrodynamics of the black holes.

Our numerical results imply that the thermodynamically preferred helical black holes (the red line in figure \ref{fig:1}) have the property that the expansion coefficient $c_h$, appearing in \eqref{asexp} and entering the definition of the energy density and the pressure in the direction of the axis of the helix (see \eqref{emtvev}), is exactly zero. The same phenomenon was also seen for the helical superconducting black holes of \cite{Donos:2012gg} and it is desirable to have a better understanding of this property.

The construction of the $D=5$ black hole solutions describing spatially modulated phases here and in \cite{Donos:2012gg} has been facilitated by the fact that 
they are static and have a Bianchi VII$_0$ symmetry. This leads to the construction of a co-homogeneity one ansatz for the $D=5$ fields and hence
solving ordinary differential equations. Moving to $D>5$ (obviously of less interest in making connections with condensed matter systems), many generalisations are possible while staying within the realm of solving ODEs. However,  constructing
$D=4$ black holes that are dual to spatially modulated phases (as in the models \cite{Donos:2011bh,Donos:2011qt}) will necessarily involve solving partial differential equations. Wile this is technically more challenging we expect to see new phenomenon.

\section*{Acknowledgements}
We thank Gary Gibbons, Gary Horowitz, Elias Kiritsis, 
Per Kraus, Don Marolf, Rob Myers, Hirosi Ooguri, Ioannis Papadimitriou,
Joe Polchinski, Theodore Tomaras and Toby Wiseman for helpful discussions. AD is supported by an EPSRC postdoctoral Fellowship.
JPG is supported by a Royal Society Wolfson Award. JPG and AD would like to thank the KITP and CCTP, respectively, for hospitality
where this work was completed. This research was supported in part by the National Science Foundation under Grant No. NSF PHY05-51164. 

\appendix

\section{More on thermodynamics}\label{moreonthermo}
Here we will expand upon the discussion of the thermodynamics that we summarised in section \ref{thermo}.
We found it illuminating to consider this issue in a slightly more general setting in which we allow for a non-normalisable fall-off in the magnetic part of the gauge field at infinity:
\begin{align}
b=\mu_{b}+\frac{c_{b}}{r^{2}}+\ldots\,.
\end{align}
We emphasise that our new helical black hole solutions all have $\mu_b=0$.
The asymptotic expansion at infinity now reads
\begin{align}\label{mubexp}
g&=r^2 \left(1-\frac{M}{r^4}+\frac{k^4 \mu_b^2+16 c_b^2+8 q^2}{24 r^6}-\frac{2 k^2 \mu_b c_b\ln r}{3 r^6}+\frac{k^4 \mu_b^2 \ln r^2}{6 r^6}+\dots\right)\nn
f&=f_0\Bigg(1+\frac{-c_h+\frac{k^2 \mu_b^2}{48}}{r^4}
-\frac{k^2 \mu_b^2 \ln r}{16 r^4}
+\frac{-49 k^4 \mu_b^2+48 k^2 \mu_b c_b-432c_b^2}{1728r^6}\nn
&\qquad\qquad\qquad\qquad\qquad\qquad\qquad\qquad+\frac{\left(- k^4 \mu_b^2+18k^2 \mu_b c_b\right) \ln r}{72 r^6}-\frac{k^4 \mu_b^2 (\ln r)^2}{16 r^6}+\dots\Bigg)\nn
 h&=r \Big(1+\frac{c_h}{r^4}+\frac{k^2 \mu_b^2 \ln r}{16 r^4}+\frac{35 k^4 \mu_b^2-96 k^2 \mu_b c_b+144 c_b^2}{1728 r^6}+\frac{\left(k^4 \mu_b^2-3 k^2 \mu_b c_b\right) \ln r}{36 r^6}\nn
 &\qquad\qquad\qquad\qquad\qquad\qquad\qquad\qquad\qquad\qquad\qquad\qquad\qquad+\frac{k^4 \mu_b^2 (\ln r)^2}{48 r^6}+\dots\Big)\nn
   \alpha&=\frac{c_\alpha}{r^4}-\frac{k^2 \mu_b^2 \ln r}{16 r^4}+\frac{576 c_\alpha k^2-59 k^4 \mu_b^2+96 k^2 \mu_b c_b-144 c_b^2}{1728 r^6}\nn
   &\qquad\qquad\qquad\qquad\qquad+\frac{\left(-7 k^4 \mu_b^2+12 k^2 \mu_b c_b\right) \ln r}{144 r^6}-\frac{k^4 \mu_b^2 (\ln r)^2}{48 r^6}+\dots\nn
 Q&=f_0\left(\frac{c_Q}{r^4}+\frac{k^2 \mu_b q-12 c_b q+3 k^2 c_Q}{36 r^6}+\frac{k^2 \mu_b q \ln r}{6 r^6}+\dots\right)\nn
 a&=f_0\left(\mu+\frac{q}{r^2}+\frac{-\gamma k^3 \mu_b^2+8 \gamma k \mu_b c_b}{32 r^4}-\frac{\gamma k^3 \mu_b^2 \ln r}{8 r^4}+\dots\right)\nn
 b&=\mu_b+\frac{c_b}{r^2}-\frac{k^2 \mu_b \ln r}{2 r^2}+\frac{-3 k^4 \mu_b+8 k^2 c_b+16 \gamma k \mu_b q}{64 r^4}-\frac{k^4 \mu_b \ln r}{16 r^4}+\dots\,.
\end{align}
The expansion at the black hole horizon is presented in \eqref{nhexp}.

We will consider the total Euclidean action, $I_{Tot}$, defined as
\begin{align}
I_{Tot}=I+I_{bndy}\,,
\end{align}
where $I=-iS$ and the (standard) Euclidean boundary action, $I_{bndy}$, is given by an integral on the boundary $r\to \infty$ \cite{D'Hoker:2009bc}:
\begin{align}
I_{bndy}=\int d\tau d^{3}x\,\sqrt{-g_{\infty}}\,\left(-2K+6-\frac{1}{4}\ln r \,F_{mn}F^{mn}+\ldots \right)\,.
\end{align}
Here $K=g^{mn}\nabla_m n_n$ is the trace of the extrinsic curvature of the boundary, where $n^m$ is an outward pointing normal vector,
and ${g_\infty}$ is the determinant of the induced metric.
The $\ln r$ term is required to remove the divergence associated with the trace anomaly
$T_m{}^m=-\frac{1}{12} F_{mn}F^{mn}$ and the ellipsis refers to a Ricci scalar term which will not be relevant for the ansatz 
and boundary conditions that we are considering.
For our ansatz we have 
\begin{align}\label{eq:bound_action}
I_{bndy}=\mathrm{vol_{3}}&\Delta\tau\,\lim_{r\to\infty}r^{2}hfg^{1/2}\Bigg[6-2g^{1/2}\,\left(\frac{2}{r}+\frac{h^{\prime}}{h}+\frac{f^{\prime}}{f} \right)-g^{-1/2}g^{\prime} \nn
&-\frac{1}{2}\ln r\left(-\frac{(a'-Qb')^2}{f^2}+\frac{e^{2\alpha}k^2b^2}{r^2h^2}+\frac{e^{-2\alpha}g(b')^2}{r^2}\right)
+\dots\Bigg]\,,
\end{align}
where $\mathrm{vol}_3=\int dx^1dx^2 dx^3$.
We next point out two equivalent ways to write the bulk part of our Euclidean action on-shell:
\begin{align}\label{eq:bulk_action}
I_{OS}&=\mathrm{vol_{3}}\Delta\tau\,\int_{r_{+}}^{\infty} dr\,\left[2rghf+\frac{r^{4}e^{2\alpha}h}{2f}QQ^{\prime}+\frac{1}{2}he^{-2\alpha}fgbb^{\prime}+\frac{1}{2f} r^{2}h\left(a^{\prime}-Qb^{\prime} \right)bQ +\frac{1}{6}k\gamma ab^{2}\right]^{\prime}\notag\\
&=\mathrm{vol_{3}}\Delta\tau\,\int_{r_{+}}^{\infty}dr\,\left[r^{2}hfg^{\prime}+2r^{2}hgf^{\prime} -\frac{h}{f}r^{4}e^{2\alpha}QQ^{\prime}-\frac{1}{f}r^{2}ha \left(a^{\prime}-Qb^{\prime} \right)-\frac{1}{3}k\gamma ab^{2}\right]^{\prime}
\end{align}
Notice that the first expression only receives contributions from the boundary at $r\to\infty$ since $g(r_+)=Q(r_+)=a(r_+)=0$, while the second expression
also receives contributions from $r=r_+$. Using the expansions at the AdS boundary \eqref{mubexp} and at the black hole horizon \eqref{nhexp}, and combining with 
\eqref{eq:bound_action} we obtain the following two equivalent expressions for the total on-shell action:
\begin{align}\label{osact2}
[I_{Tot}]_{OS}&=\mathrm{vol_{3}}\frac{1}{T}[-M-\mu_b\,c_b-\frac{1}{12}\mu_b^{2}k^{2}+\frac{1}{6}\mu\mu_b^{2}k\gamma] \notag\\
&=\mathrm{vol_{3}}\frac{1}{T}\left[3M+8c_{h} +2\mu q-Ts-\frac{1}{8}\mu_b^{2}k^{2}-\frac{1}{3}\mu \mu_b^{2}k\gamma\right]\,.
\end{align}

A variation of the bulk action $I$ gives equations of motion and boundary terms. 
Thus an on-shell variation only gets contributions from the boundary. We hold $k$ fixed in these
variations and then we only get contributions at $r\to\infty$. 
Combining this with an on-shell variation of the boundary action
$I_{bndy}$ and using the asymptotic expansion \eqref{mubexp} we eventually obtain
\begin{align}
[\delta I_{Tot}]_{OS}=&\mathrm{vol_{3}}\Delta\tau\,\Bigg[\left(8c_{h}+3M+2\mu q-\frac{1}{8}\mu_b^{2}k^{2}-\frac{1}{3}\mu\mu_b^{2}k\gamma \right)\delta f_{0}+f_{0}\left(2q-\frac{1}{3}\mu_b^{2}k\gamma \right)\,\delta\mu \notag\\
& \qquad\qquad\qquad\qquad +f_{0}\left(-2c_b-\frac{1}{2}\mu_bk^{2}+\frac{1}{3}\mu\mu_bk\gamma \right)\,\delta\mu_b\Bigg]\,.
\end{align}
In this variation we are holding $\Delta\tau$ fixed and hence $\Delta\tau\delta f_0=-T^{-2}\delta T$.
We next define the potential $W$, and a corresponding density $w$, for the grand canonical ensemble via $W=T[I_{Tot}]_{OS}=w\mathrm{vol}_3$. We deduce that $w=w(T,\mu,\mu_b)$ with
\begin{align}\label{varwgen}
\delta w=-s \delta T+\left(2q-\frac{1}{3}\mu_b^{2}k\gamma \right)\,\delta\mu-\left[2c_b+\frac{1}{2}\mu_bk^{2}-\frac{1}{3}\mu\mu_bk\gamma \right]\,\delta\mu_b\,.
\end{align}

We now compute the expectation value of the stress tensor and the current. For the former we have
\cite{D'Hoker:2009bc}
\begin{align}\label{emdefb}
\langle T_{mn}\rangle=\lim_{r\to\infty}r^2\left[-2K_{mn}+2(K-3)(g_\infty)_{mn}+\left(F_{m}{}^{p}F_{np}-\frac{1}{4}(g_\infty)_{mn} F_{pq}F^{pq}\right)\,\ln r +\dots\right]\,.
\end{align}
Using our expansion at the AdS boundary \eqref{mubexp} we obtain
\begin{align}\label{emtvev2}
\langle T_{tt}\rangle&=\,f_{0}^{2}\,\left(3M+8c_{h}-\frac{1}{8}\,\mu_b^{2}k^{2}\right)\notag\\
\langle T_{tx_{2}}\rangle&=\,4f_0c_{Q}\,\cos\left(k x_{1}\right)\notag\\
\langle T_{tx_{3}}\rangle&=-4f_0c_{Q}\,\sin\left(k x_{1}\right)\notag\\
\langle T_{x_{1}x_{1}}\rangle&=M+8c_{h}-\frac{7}{24}\,\mu_b^{2}k^{2}\notag\\
\langle T_{x_{2}x_{2}}\rangle&=M +\left(8c_{\alpha}+\frac{1}{8}\,\mu_b^{2}k^{2}\right)\,\cos\left(2kx_{1} \right) \notag\\
\langle T_{x_{3}x_{3}}\rangle&=M -\left(8c_{\alpha}+\frac{1}{8}\,\mu_b^{2}k^{2}\right)\,\cos\left(2kx_{1} \right) \notag\\
\langle T_{x_{2}x_{3}}\rangle&=-\left(8c_{\alpha}+\frac{1}{8}\,\mu_b^{2}k^{2}\right)\,\sin\left(2kx_{1}\right)\,.
\end{align}
Observe that $\langle T_m{}^m\rangle=-\mu_b^2k^2/6 =-\frac{r^2}{12} F_{mn}F^{mn}$, as expected (here we are raising indices with
$g_\infty^{mn}$ and the $r^2$ factor appears because it also appears in \eqref{emdefb}) .
Setting $f_0=1$ we see that the energy density is given by
\begin{align}
\varepsilon=3M+8c_{h}-\frac{1}{8}\,\mu_b^{2}k^{2}\,.
\end{align}
Observe that the equality of the two expression in \eqref{osact2} imply the Smarr type formula
\begin{align}
\tfrac{4}{3}\varepsilon=sT-2\mu q+\tfrac{8}{3}c_h-\mu_bc_b-\tfrac{1}{8}\mu^2_bk^2+\tfrac{1}{2}\mu\mu_b^2k\gamma\,.
\end{align}
If we define the average pressure $\bar p=(\langle T_{x_1x_1}\rangle+\langle T_{x_1x_1}\rangle+\langle T_{x_1x_1}\rangle)/3$, this can also be written in the form 
\begin{align}\label{smarrn}
\varepsilon+\bar p=sT-2\mu q+\tfrac{8}{3}c_h-\mu_bc_b-\tfrac{13}{72}\mu^2_bk^2+\tfrac{1}{2}\mu\mu_b^2k\gamma\,.
\end{align}
We next calculate the expectation value of the current. The relevant terms are given by\cite{D'Hoker:2009bc}
\begin{align}\label{eq:current2}
\langle J_{m}\rangle=\lim_{r\to\infty}\,\left[r^3 F_{rm}-\frac{1}{6}\gamma\, \epsilon_{m}{}^{npq}A_{n}F_{pq}+{\nabla}_{n}F^{n}{}_{m}\,\ln r+\dots\right]\,.
\end{align}
where ${\nabla}$ is the Levi-Civita covariant derivative with respect to the boundary metric $g_\infty$.
Using the expansion \eqref{mubexp} we obtain 
\begin{align}\label{vevofJ2}
\langle J_{t}\rangle&=-2f_0q+\frac{1}{3}\mu_b^{2}k\gamma\notag\\
\langle J_{x_1}\rangle&=0\notag\\
\langle J_{x_2}\rangle&=-\left(2c_b+\frac{1}{2}\mu_bk^{2}-\frac{1}{3}\mu \mu_bk\gamma \right)\cos(kx_1)\nn
\langle J_{x_3}\rangle&=\left(2c_b+\frac{1}{2}\mu_bk^{2}-\frac{1}{3}\mu \mu_bk\gamma \right)\sin(kx_1)\,.
\end{align}
In terms of the current, the Smarr formula \eqref{smarrn} can be written in the form
\begin{align}\label{smarrnl}
\varepsilon+\bar p=sT+\mu\langle J_t\rangle+\tfrac{8}{3}c_h
+\tfrac{1}{2}\mu_b\left(\langle J_{x_2}\rangle\cos k x_1-\langle J_{x_3}\rangle \sin kx_1    \right)
+\tfrac{5}{72}\mu^2_bk^2\,.
\end{align}

As we discussed in section 2.3 we view $k$ as labelling a particular mode and hence should not be varied. 
To amplify this point, 
it is useful to refer back to our ansatz \eqref{eq:ansatz} and define the $x_1$ dependent variation $[\delta A(k)]_m$ via
\begin{align}
\left[\delta A(k)\right]_{t}&=\delta\mu\nn
\left[\delta A(k)\right]_{x_1}&=0\nn
\left[\delta A(k)\right]_{x_2}&=\delta\mu_b\,\cos(kx_{1})\nn
\left[\delta A(k)\right]_{x_3}&=-\delta\mu_b\,\sin(kx_{1})\,.
\end{align}
We then find that the first law \eqref{varwgen} can be written in the form
\begin{align}\label{firstlaw2}
\delta W=\int dx_1dx_2dx_3\Big( -s\,\delta T+\langle J^{m}(-k)\rangle \left[\delta A(k)\right]_m\Big)\,,
\end{align}
where $\langle J^{m}(-k)\rangle$ is given in \eqref{vevofJ2} and 
we note the integrand is actually independent of $x_1$.
In particular, we interpret $\delta\mu_b$ as parametrising a specific mode, labelled by $k$, of a non-normalisable deformation of the gauge-field, $\delta A(k)$.

\subsection{Another perspective on \eqref{eq:deltak}}

Let us consider a variation of the gauge field part of the bulk action \eqref{eq:lagemcs}.
This gives the boundary term
\begin{align}
\delta S_{gauge}=\int_{\partial M} \left[\ast {F}+\frac{\gamma}{3}{A}\wedge {F} \right]\wedge \delta {A}
\end{align}
Here we would like to focus on variations of wave-number $k$ which give rise to a contribution at the boundary
of the non-compact $x_1$ direction. Given our ansatz \eqref{eq:ansatz} we have
\begin{align}
\delta A=-b\,x_{1}\,\omega_{3}\, \delta k
\end{align}
We take the boundary to be at $x_1=\pm L/2$ and then take $L\to\infty$. A calculation shows that the relevant part of
the integrand is
\begin{align}
\left[\ast F+\frac{\gamma}{3}A\wedge F \right]\wedge \delta A=
-dt\wedge dx_2\wedge dx_3\wedge dr\left\{\frac{ke^{2\alpha}fb^2}{h}-\frac{\gamma}{3}b(ba'-ab')\right\}x_1\delta k+\dots
\end{align}
After evaluating this at $x_1=\pm L/2$ and then dividing by $L$
in order to find the variation of the density, 
we are led to the third and fourth terms of \eqref{eq:deltak}
(taking into account a minus sign since here we are looking at the Minkowski signature space-time).

If we now vary the Einstein-Hilbert term we get the boundary term
\begin{align}
\delta S_{EH}= \int_{\partial M}\,\sqrt{-\gamma}\,n^{\mu}\,\left(\nabla^{\nu}\delta g_{\mu\nu}-g^{\nu\lambda}\nabla_{\mu}\delta g_{\nu\lambda} \right)
\end{align}
where $\gamma$ is the induced metric on the boundary $\partial M$ and for the boundary components defined by $x_{1}=\pm L/2$ we have that the unit normal vector is $n=h^{-1}\,\partial_{x_1}$. Substituting a variation of the metric obtained by varying $k$ in our ansatz \eqref{eq:ansatz}, 
we obtain contributions at $x_1=\pm \infty$ which lead to the first and second terms in our formula \eqref{eq:deltak}, again up to a minus sign.

This calculation show that variations of $k$ in our ansatz \eqref{eq:ansatz} are associated with boundary contributions to the variation of the action 
at $x_1=\pm \infty$. Since we do
not want to modify boundary conditions at $x_1=\pm \infty$, $k$ is a parameter to be held fixed in obtaining the relevant equations of motion. 

\section{Linearised Analysis}\label{linal}
We summarise the analysis of linearised perturbations about the AdS-RN black brane solution  
considered in \cite{Nakamura:2009tf} in the language of this paper.
Specifically, we consider the perturbation
\begin{align}
Q\rightarrow \epsilon\, Q,\qquad 
b\rightarrow \epsilon\,b
\end{align}
with $h=\alpha=0$ in \eqref{eq:ansatz} around the AdS-RN black brane solution \eqref{eq:ansatzrn}, for small $\epsilon$.
At first order in $\epsilon$ we obtain two coupled ODEs, linear in $Q$ and $b$, given by
\begin{align}\label{linodes}
r^{-5}\left(r^{5}Q^{\prime}\right)^{\prime}-k^{2}r^{-2}g^{-1}\,Q+r^{-2}\,a^{\prime}b^{\prime}=0\,,\nn
r^{-1}g^{-1}\,\left(rg b^{\prime} \right)^{\prime}-k^{2}r^{-2}g^{-1}\,b+r^{2}g^{-1}a^{\prime}\,Q^{\prime}+kr^{-1}g^{-1}\gamma a^{\prime}\,b=0\,,
\end{align}
with $g$ and $a$ as given in \eqref{gaval}. To make contact with equation (4.17) of \cite{Nakamura:2009tf} one should make
the following identifications: $u=r_+/r$, $\psi(u)=-\sqrt{3} r^3 Q'(r)/r_+$, $b(r)=\phi(u)$, $q=\mu/(r_+\sqrt{3})$, $k_{there}=k_{here}/r_+$, $\alpha=\gamma/4$, $q=\mu/(r_+\sqrt{3})$.

In order to find a normalisable static linearised mode of interest we should impose the following boundary conditions.
At the horizon, $r\to r_+$, we demand that
\begin{align}
&Q= Q_{(+)}\left(r-r_{+}\right)+\ldots\,,\nn
&b=b_{+}+\ldots\,.
\end{align}
We are only interested in deformations of the CFT given by the temperature $T$ and the chemical potential $\mu$. Hence as $r\to \infty$
we demand that 
\begin{align}
Q=\frac{c_{Q}}{r^{4}}+\ldots\,,\nn
b=\frac{c_{b}}{r^{2}}+\ldots\,.
\end{align}
A solution to \ref{linodes} is specified by four integration constants and hence, for a given $T,\mu$, we expect a unique solution (if any).
By numerically solving \eqref{linodes} we find solutions that are summarised in figure \ref{fig:lin}, for the special value $\gamma=1.7$ (and also $\mu=1$), 
which agrees well with figure 2 of \cite{Nakamura:2009tf}.

\bibliographystyle{utphys}
\bibliography{helical}{}

\end{document}